\documentclass[aps,letterpaper,twocolumn,preprintnumbers,floatfix,superscriptaddress]{revtex4}

\usepackage{graphicx}
\usepackage{dcolumn}
\usepackage{bm}
\usepackage{epsfig}
\usepackage{gensymb}
\usepackage{mathrsfs}
\usepackage{amsmath}
\usepackage{amssymb}
\usepackage{graphicx,bm}
\usepackage{natbib}
\usepackage{slashed}
\usepackage[makeroom]{cancel}
\usepackage{dsfont}
\usepackage{longtable}
\usepackage{multirow}
\usepackage{youngtab}
\usepackage{young}
\usepackage{tikz}
 \usepackage[titletoc]{appendix}

\def\fun#1#2{\lower3.6pt\vbox{\baselineskip0pt\lineskip.9pt
  \ialign{$\mathsurround=0pt#1\hfil##\hfil$\crcr#2\crcr\sim\crcr}}}

\usepackage{tikz}
\usetikzlibrary{decorations.pathmorphing,decorations.markings,trees,shapes}

\def\lsim{\mathrel{\rlap{\raise 2.5pt \hbox{$<$}}\lower 2.5pt\hbox{$\sim$}}}
\def\gsim{\mathrel{\rlap{\raise 2.5pt \hbox{$>$}}\lower 2.5pt\hbox{$\sim$}}}

\input epsf

\newcommand{\blue}[1]{\textcolor{blue}{#1}}
\usepackage{color}

\newcommand{\comment}[1]{}

\newcommand{\be}{\begin{equation}}
\newcommand{\ee}{\end{equation}}
\newcommand{\bea}{\begin{eqnarray}}
\newcommand{\eea}{\end{eqnarray}}

\newcommand{\vev}[1]{\langle #1 \rangle}
\newcommand{\ket}[1]{| #1 \rangle}
\newcommand{\bra}[1]{\langle #1 |}
\newcommand{\antiket}[1]{| #1 ]}

\newcommand{\eq}[1]{\begin{equation}\begin{split} #1 \end{split}\end{equation}}

\renewcommand\arraystretch{2}

\begin{document}

\title{Constructing on-shell operator basis for all masses and spins}

\author{Zi-Yu Dong}
\affiliation{CAS Key Laboratory of Theoretical Physics, Institute of Theoretical Physics,
Chinese Academy of Sciences, Beijing 100190, China.}
\affiliation{School of Physical Sciences, University of Chinese Academy of Sciences, Beijing 100190, P. R. China.}
\author{Teng Ma}
\affiliation{Physics Department, Technion -- Israel Institute of Technology, Haifa 3200003, Israel}
\author{Jing Shu}
 \affiliation{CAS Key Laboratory of Theoretical Physics, Institute of Theoretical Physics,
Chinese Academy of Sciences, Beijing 100190, China.}
\affiliation{School of Physical Sciences, University of Chinese Academy of Sciences, Beijing 100190, P. R. China.}
    \affiliation{CAS Center for Excellence in Particle Physics, Beijing 100049, China}
    \affiliation{School of Fundamental Physics and Mathematical Sciences, Hangzhou Institute for Advanced Study, University of Chinese Academy of Sciences, Hangzhou 310024, China}
    \affiliation{Center for High Energy Physics, Peking University, Beijing 100871, China}
    \affiliation{International Center for Theoretical Physics Asia-Pacific, Beijing/Hanzhou, China}

\begin{abstract}
We first propose a general method to construct the complete set of on-shell operator bases involving massive particles with any spins. To incorporate the non-abelian little groups of massive particles, the on-shell scattering amplitude basis should be factorized into two parts: one is charged, and the other one is neutral under little groups of massive particles. The complete set of these two parts can be systematically constructed by choosing some specific Young diagrams of Lorentz subgroup and global symmetry $U(N)$ respectively ($N$ is the number of external particles), without the equation of motion and integration by part redundancy. Thus the complete massive amplitude bases without any redundancies can be obtained by combining these two complete sets. Some examples are presented to explicitly demonstrate this method.  This method is applicable for constructing amplitude bases involving identical particles, and all the bases can be constructed automatically by computer programs based on it.

\end{abstract}

\pacs{xxx}

\maketitle

\section{Introduction}
The standard model (SM) successfully describes particle physics up to TeV scale. But it still can not explain a lot of profound puzzles in particle physics, such as Higgs naturalness~\cite{Wilson:1970ag,tHooft:1979rat}, dark matter~\cite{Zwicky:1933gu,Rubin:1970zza}, neutrino mass~\cite{Davis:1968cp,Pontecorvo:1957cp}, and so on. These unsolved puzzles strongly suggest that SM is not complete and new physics (NP) should be introduced. But so far any physics beyond SM is not observed at LHC and other detections, which indicates that the scale of NP may be too high to be within the reach of current experiments on the ground.

For the high energy scale NP, effective field theory (EFT) is the critical tool to study its lower energy effects. The imprints of the NP can be detected by measuring higher dimensional operators at lower energy scale experiments, even the null NP signals from direct detections. The completeness of the EFT operator bases is essential to capture the full imprints of NP. So how to construct the complete bases systematically without redundancies from the equation of motion (EOM) and integration by part (IBP) is critical for EFT formulation. In traditional field theory, it is still an unsolved issue (the operators are usually constructed by hands), especially in dealing with EOM and IBP, though the number of the bases can be systematically counted through Hilbert series technique~\cite{Lehman:2015via,Lehman:2015coa,Henning:2015alf,Henning:2017fpj}.

Recently it was found that the on-shell scattering amplitude is very efficient to deal with EFT without referring to the Lagrangian, such as calculating SMEFT running~\cite{Bern:2020ikv,Jiang:2020mhe,EliasMiro:2020tdv,Baratella:2020lzz}, deriving selecting rules~\cite{Cheung:2015aba,Jiang:2020sdh}, and constructing some special EFTs through soft limit~\cite{Cheung:2014dqa,Cheung:2016drk,Low:2014nga,Low:2014oga}. One of the most appealing features of the on-shell method is that it can efficiently construct the complete EFT bases, called amplitude bases (an independent unfactorizable scattering amplitude corresponds to an independent operator)~\cite{Elvang:2010jv,Shadmi:2018xan,Ma:2019gtx}. For the general massless EFT, the on-shell method can automatically eliminate EOM redundancy, and the complete amplitude bases can be systematically constructed without IBP redundancy through Young tableaus of the global symmetry of massless spinors~\cite{Henning:2019enq,Henning:2019mcv} (its extension in constructing the amplitude bases involving identical particles is discussed in~\cite{Li:2020gnx}). However, this method is only applicable for EFT of massless particles, and generally, the SM particle mass is always relevant in lower energy scale measurements. So this method is not helpful in  lower energy phenomenology study. It is troublesome when applying these bases in practical calculations: we should first translate the amplitude bases into the field operators and then use them to study lower energy phenomenology at the electroweak symmetry breaking (EWSB) phase as did in traditional SMEFT. Besides, it was found that many operators in the massless SMEFT contribute to the same physics process after EWSB, so the massless SMEFT is redundant to describe some specific NP effects at the EWSB phase. For example, there are seven independent operators fully describing three massive gauge bosons effective interactions while there are infinite operators in massless SMEFT that contribute to them, such as the type of operators $\left(W_{\mu\nu}^a \right)^3 H^{n} H^{\dagger n}$ with $n$ being any positive integer. These shortcomings indicate that massive EFT is more concise to describe NP effects at a lower energy scale than massless EFT.

To formulate the massive EFT, the complete operator bases should be systematically constructed. However, so far there is no principle to construct the complete massive bases even there are some primary explorations done in~\cite{Durieux:2019eor,Durieux:2020gip,Falkowski:2020fsu,Murphy:2020cly}. In this paper, we first propose a method to systematically construct the on-shell amplitude bases involving massive particles with any spin based on group theory. Since the little group (LG) of a massive particle is non-abelian $SU(2)$, we first split the massive amplitude basis into two parts: the massive LG tensor structure (MLGTS), which is the holomorphic function of massive right-handed spinors and only charged under massive LG, and massive LG neutral structure (MLGNS) which is only charged under massless LG. Then construct the complete set of these two parts separately and finally combine the two complete sets to get the complete massive amplitude basis.

We find MLGTSs can be classified by Lorentz subgroup $SU(2)_r$ representations and thus can be completely constructed by finding all the  $SU(2)_r$ representations allowed by massive LG (their representations are correlated). While MLGNSs suffer from EOM and IBP redundancy. Since the massless spinors can automatically eliminate the EOM redundancy, we can first construct MLGNSs at massless limit and then one to one map their massless limit into MLGNSs without worrying about EOM redundancy. So the only job to constructing MLGNS is to completely construct its massless limit structure. Since the massless structures also suffer from IBP redundancy, we can embed the spinors associated with $N$ external particles into $U(N)$ representation. Then the complete set of the massless limit of MLGNSs can be systematically constructed by some special $U(N)$ Semi-standard Young Tableaus (SSYTs), which can automatically eliminate IBP redundancy~\cite{Henning:2019enq}. We prove that the complete massive amplitude bases without EOM and IBP redundancy can be obtained by this method. We give some examples to explicitly construct the massive bases from $SU(2)_r$ Young diagrams (YDs) and $U(N)$ SSYTs, and also briefly discuss how to construct massive amplitude bases involving identical particles.

The structure of this paper is organized as follows. In Sec.~\ref{sec:spinor}, we briefly introduce the spinors and LG symmetry of on-shell scattering amplitudes.  In Sec.~\ref{sec:basis}, we demonstrate in detail how to construct systematically massive amplitude bases. In Sec.~\ref{sec:example}, some examples for constructing massive bases are given. In Sec.~\ref{sec:identical}, we briefly explain how to deal with identical particles. We conclude in Sec.~\ref{sec:conclusion}. The appendices contain the proof of massive amplitude bases independence, explicit massive bases expressions, and an example of constructing amplitude bases involving identical particles.

\section{Massless and Massive spinors}
\label{sec:spinor}
In this section, we briefly introduce the massive and massless spinors and the basic property of on-shell scattering amplitudes.

For a particle-$i$, its momentum can be written as a product of two spinors~\cite{Witten:2003nn,Arkani-Hamed:2017jhn},
\bea
    (p_i)_{\dot{\alpha}\alpha}\equiv (p_i)_\mu(\sigma^\mu)_{\dot{\alpha}\alpha}
    =|i^I]_{\dot\alpha}\langle i_I|_{\alpha},
\eea
and similarly for $p_i^{\alpha\dot{\alpha}}\equiv (p_i)_\mu(\bar{\sigma}^\mu)^{\alpha\dot{\alpha}}$, where $\sigma^\mu (\bar{\sigma}^\mu) \equiv \{1, (-)\sigma^i \}$ with $ \sigma^i$ is Pauli-matrices, the right-handed spinor $\antiket{i^I}_{\dot \alpha}$ and left-handed spinor $\ket{i^I}_{\alpha}$  is in the fundamental representation of Lorentz subgroup $SU(2)_r$ and $SU(2)_l$ ($SO(3,1) \simeq SU(2)_l \times SU(2)_r$),        $I$ is the index of LG $SU(2)_i$ or $U(1)_i$ for massive or massless particle-$i$. For spinors of massless momentum, the index $I$ is trivial and can be neglected. The massive right/left handed spinor $\antiket{i^I}_{\dot{\alpha}}$/$\ket{i^I}_{\alpha}$ is in the fundamental representation of its LG $SU(2)_i$ and massless right/left handed spinor $\antiket{i}_{\dot{\alpha}}$/$\ket{i}_{\alpha}$ takes $+/-$ unite charge of its LG $U(1)_i$.
Two spinors can form a Lorentz scalar through contracting their Lorentz index with two-index Levi-Civita tensor $\epsilon^{\dot \alpha  \dot \beta}$ ($\epsilon^{\alpha \beta}$) and we can introduce the square spinor bracket $[ij]$  and angle spinor bracket $\vev{ij}$ to denote this spinor product,
\bea
[ij]^{IJ} \equiv \epsilon^{\dot  \alpha \dot  \beta} \antiket{i^I}_{\dot\beta} \antiket{j^J}_{\dot\alpha}\,, \quad
\vev{ij}^{IJ} \equiv  \epsilon^{\alpha \beta} \ket{i^I}_{\beta} \antiket{j^J}_{ \alpha}\,.
\eea
For massive spinor, the left- and right-handed spinor can be related with each other through EOM,
\bea
    &p_i|i^I]=m_i|i^I\rangle\,,\quad p_i|i^I\rangle=-m_i|i^I]\,.
\eea
The on-shell scattering amplitudes can be expressed generally in terms of the angle and square brackets of external momenta together with gauge symmetry structures. If an external particle-$i$ is massive with spin $s_i$, the scattering amplitude should be in the $2s_i$ indices symmetric  representation of LG $SU(2)_i$ (i.e. $2s_i+1$ dimension representation)  while it should take $2 h_i$ charge of massless LG $U(1)_i$ if the external particle-$i$ is massless with helicity $h_i$. For example, the amplitude with an external massive vector particle-$i$ should transform under LG $SU(2)_i$ as~\cite{Arkani-Hamed:2017jhn}
\bea
&& \mathcal{M}^{\left \{I_1,I_{2} \right\}}\left(w_{I I^\prime}^i \antiket{i^{I^\prime}}, w_{I I^\prime}^i \ket{i^{I^\prime}},\ldots \right) \nonumber \\
&&=  w_{I_1 I^\prime_1}^i  w_{I_2 I^\prime_2}^i  \mathcal{M}^{\{I^\prime_1,I^\prime_{2}\}}\left(\antiket{i^{I^\prime}},  \ket{i^{I^\prime}},\ldots \right) \,,
 \eea
where $w^i$ is LG $SU(2)_i$ element and the superscript bracket  $\left \{I^i_1,\dots,I^i_{2s_i} \right\}$ means that these $2s_i$  indices of $SU(2)_i$ should be totally symmetric.

\section{Constructing Amplitude Basis}
\label{sec:basis}
According to the LG transformation of the scattering amplitude, its general structure can be factorized into two parts: MLGTS and MLGNS. This general structure indicates that, to get the complete amplitude bases, we can separately construct the complete set of tensor structures and the corresponding neutral structures. We find that  MLGTS and MLGNS can be constructed completely through Lorentz subgroup $SU(2)_r$ and a $U(N)$ global symmetry respectively ($N$ is the number of external particles)  without EOM and IBP redundancy. In this section, we will discuss in detail how to construct these two parts completely in a systematic way.

\subsection{Massive LG Tensor Structures}
As said before, the scattering amplitudes of $m$ massive and $n$ massless particles can be factorized into MLGTS and MLGNS,
\bea
\mathcal{M}^I_{m,n} =\sum_{\{\dot \alpha \}}  \mathcal{A}^{ I}_{\{\dot \alpha \} } \left(\left \{\epsilon_{s_i} \right \} \right) G^{\{\dot\alpha \}}\left(\antiket{j},\ket{j},p_{i} \right),
\eea
where $\epsilon_{s_i} \equiv  \antiket{i}^{\{ I_{1}}_{\dot \alpha_1},\ldots, \antiket{i}^{I_{2s_i} \}}_{\dot\alpha_{2s_i}} $ is the polarization  tensor of massive particle-$i$ with spin $s_i$, whose quantum number under $SU(2)_i \otimes SU(2)_r$ is $(2s_i+1,2s_i+1)$, $I$ generally denotes the   LG indices and $\{\dot\alpha\}$ collectively denote Lorentz indices. Since the massive left and right handed spinors are related by EOM, $ \ket{i^I}  = p_i  \antiket{i^I}/m_i$, the MLGTS $\mathcal{A}^{I}$ is required to be the holomorphic function of right-handed massive spinors $ \antiket{i^I}$ without losing generality, which indicates that $\mathcal{A}^{I}$ totally contain $2s_i$ massive spinor $\antiket{i^I}$ associated with particle-$i$ with spin $s_i$. Moreover, since $\mathcal{A}^I_{\{\dot \alpha \} }$ must be in dim-$(2s_i+1)$ representation of LG $SU(2)_i$ and the permutation symmetry in $SU(2)_i$ indices should transfer into Lorentz $SU(2)_r$ indices of spinors $\antiket{i^I}_{\dot\alpha}$, all the same spinors $ \antiket{i^I}_{\dot\alpha}$ in $\mathcal{A}^I_{\{ \dot \alpha \} }$ should be in the $(2s_i+1, 2s_i+1)$ representation of $SU(2)_i \otimes SU(2)_r $, which indicates that MLGTS is only the linear function of polarization tensor $\epsilon_{s_i}$. This property indicates that $\mathcal{A}^I_{\{ \dot \alpha \} }$ is free of EOM and it is also free of IBP redundancy because it does not contain momentum $p_i$.  MLGNS  $G( \antiket{j},\ket{j}, p_{i})$ is only charged under massless LG and neutral under massive LG so it is the function of massless spinors $\antiket{j}$ or $\ket{j}$ and massive momentum $p_{i}$.
For the same scattering process, different MLGTSs are in the same massive LG $\otimes^m_{i=1}SU(2)_i$ representation so massive LG can not classify MLGTS. But since $\mathcal{A}^{I}$ is also Lorentz subgroup $SU(2)_r$ tensor, it can be classified by $SU(2)_r$ representation. Similar to constructing the wave function of composite resonances in QCD via global $SU(3)$ symmetry SSYTs, the tensor structure $\mathcal{A}^{I}$ can be completely constructed through finding all of its possible $SU(2)_r$ representation via YD. Next we will discuss in detail how to systematically construct the complete MLGTS.

As said before, the quantum number of massive left-handed spinor $ \antiket{i^I}_{\dot \alpha}$ under $SU(2)_i \otimes SU(2)_r$ is $(2,2)$, represented by YD as
\bea
 \antiket{i^I}_{\dot \alpha} = \Yvcentermath1  {\young(i)}_i \otimes \young(i)_r,
\eea
where we use $i$ filled in the box to label YD of spinor $\antiket{i^I}_{\dot \alpha}$ and the subscript $i$ or $r$  to label $SU(2)_i$ or $SU(2)_r$ YD (notice that the LG of different spinors is different). Thus a  YD of massive LGs and $SU(2)_r$ correspond to an independent holomorphic function of right-handed massive spinors and this function can be written down according to the group indices permutation symmetry of this YD.  For example, two massive right-handed spinors product can be read from the following YD,
\bea \label{eq:spinor_product1}
  \Yvcentermath1  {\young(i)}_i \otimes { \young(j) }_j \otimes {\young(i,j)}_r  = \antiket{i^I}_{\dot\alpha}  \antiket{j^J}_{\dot\beta} - \antiket{i^I}_{\dot\beta}  \antiket{j^J}_{\dot\alpha} =[i^Ij^J].
   \eea
 The massive polarization tensor $\epsilon_{s_i}$ in the $(2s_i +1,2s_i+1)$ representation of $SU(2)_i \otimes SU(2)_r$ can be read from the direct product of two dim-$(2s_i +1)$ YDs with one row and $2s_i$ columns,

\bea \label{eq:tensor_YD}
{\underbrace{\Yvcentermath1 { \young(i)\cdots \young(i)}_i}_{(2s_i)}}  \otimes \underbrace{\Yvcentermath1 {\young(i)\cdots\young(i)}_r}_{(2s_i)}=|i^{\{I_1}]_{\dot{\alpha}_1}
\!\cdots|i^{I_{2s_i}\}}]_{\dot{\alpha}_{2s_i}}  = \epsilon_{s_i}\,.
\eea
As said before, the correlation between the permutation symmetry of $SU(2)_r$ and $SU(2)_i$ indices of all the spinors $\antiket{i^I}$ in MLGTS requires that $\mathcal{A}^I_{\{ \dot\alpha \} }$ must be the linear function of polarization tensor $\epsilon_{s_i}$. Since the LG indices of these $m$ polarization tensors do not contract with each other,  only the different contraction pattern of their $SU(2)_r$ indices can generate different $\mathcal{A}^I_{\{ \dot\alpha \} }$. So MLGTS can be classified by Lorentz $SU(2)_r$ irreducible representations of these polarization tensors even different $\mathcal{A}^I_{\{ \dot\alpha \} }$ are in the same representation of LG $\otimes^m_{i=1}SU(2)_i$. So, to construct the complete MLGTS $\mathcal{A}^I_{\{ \dot\alpha \} }$,  we just use YD to systematically find all the irreducible $SU(2)_r$ representations decomposed from the outer product of these $m$ polarization tensor YDs in Eq.~(\ref{eq:tensor_YD}) based on Littlewood-Richardson Rule and then read out these structures from YDs according to $SU(2)_r$ indices permutation symmetry. With this method, we can systematically construct all the independent MLGTSs.

Next we explicitly demonstrate how to use YD to construct MLGTS.
Take $4$-point massive amplitude basis for fermion-fermion-vector-scalar $\psi \psi^\prime Z h$ as an example.
According to Eq.~(\ref{eq:tensor_YD}), polarization tensors associated with these four external legs are in the following representations of $SU(2)_r$ (we neglect the subscript $r$ of $SU(2)_r$ YDs)
\bea
     \Yvcentermath1 \psi \sim \young(1)\quad \psi^\prime  \sim\young(2)
     \Yvcentermath1 \quad  Z \sim\young(33)\quad h \sim\bullet,
\eea
where the number in the box is used to distinguish the $SU(2)_r$ indices of different polarization tensors and the bullet $\bullet$ represents the trivial $SU(2)_r$ representation. Then we can reduce the outer product of these 4 $SU(2)_r$ tensors to the irreducible representations by Littlewood-Richardson Rule and get four representations of $SU(2)_r$ (outer product between YDs are denoted as  $\times$),
\bea
  &&  \Yvcentermath1 \young(1)\times\young(2)\times\young(33)\times\bullet \nonumber \\
      &&= \Yvcentermath1 \young(12,33)\oplus\young(123,3)\oplus\young(133,2)\oplus\young(1233).
\eea
Notice that in this work we follow the following convention of order rule in filling $SU(2)_r$ YD:
\bea
\# 1 \mbox{s} \le \# 2\mbox{s} \le \ldots \le \# m \mbox{s},
\eea
so the higher spin particle should be labeled by bigger number (in this example we label the spinors of $Z$ by the largest number among the four legs). Then we can read out the expression of MLGTS according to permutation symmetry of $SU(2)_r$ indices (the indices of different polarization tensor are labeled by different number) from above YDs, as did in Eq.~(\ref{eq:spinor_product1}). Taking the first YD as an example, the corresponding MLGTS is given by
\bea
&&\mathcal{A}^I_{[2,2]} \equiv \Yvcentermath1 \young(12,33) \nonumber \\
&&= (\antiket{1^I}_{\dot\alpha} \antiket{2^J}_{\dot\beta} \antiket{3^{K_1}}_{\dot\gamma_1} \antiket{3^{K_2}}_{\dot\gamma_2}  +\mbox{perms in}\, SU(2)_r\, \mbox{indices} )\nonumber \\
&&=[1^I3^{\{K_1}][2^J3^{K_2\}}].
\eea

Finally we can easily prove that the amplitude bases with tensor structures in different $SU(2)_r$ representations must be independent. The amplitude basis with a MLGTS in the $SU(2)_r$ representation denoted by $[\lambda]$ can be parametrized as $\mathcal{A}_{[\lambda]}^I\cdot G(\ket{j}, \antiket{j},p_{i})$, where the dot product means $\dot \alpha$ indices contraction. If the MLGTS of two amplitude bases are in different YD representation, these two bases should be two independent massive LG tensors. So, no matter what the $SU(2)_r$ YD of $G$ is, they must be two independent $SU(2)_r$ singlet states.

We use $SU(2)_r$ YD to construct the complete set of MLGTS and find that the amplitude bases with different $SU(2)_r$ MLGTS are independent. For the amplitude bases with the same MLGTS $\mathcal{A}_{[\lambda]}^I$, we can also use group theory to systematically construct the complete independent MLGNS $G$ by embedding the spinors associated with $N \equiv m+n$ external legs into a global symmetry $U(N)$ representation~\cite{Henning:2019enq,Henning:2019mcv}. Similar to above discussion, $G$ can be classified by $U(N)$ representation, each $G$ correspond to a basis of a special $U(N)$ representation so all the $G$ are independent. Also, this method can systematically get rid of the redundancy from EOM and IBP. In the next subsection, we will discuss these issues in detail.

\subsection{Massive LG neutral Structures}
 For the on-shell massless SMEFT, the amplitude bases can automatically
 get rid of EOM redundancy because any basis containing this redundancy is null ($p_j \antiket{j} =0$). So the only issue for constructing massless amplitude bases is to remove IBP redundancy, which is systematically solved by YD method~\cite{Henning:2019enq,Henning:2019mcv}. However the situation is different when constructing on-shell massive SMEFT.
Since the EOM of massive spinor is not as trivial as massless spinor, $p_i \antiket{i^I} =m_i  \ket{i^I}$, a redundant on-shell basis can not only be expressed as  the combination of the independent bases with the same dimension through IBP (i.e. momentum conservation) but also the combination of lower dimensional bases multiplied by the mass factors through EOM. So both redundancy from EOM and IBP should be removed
in constructing on-shell massive EFT. Since MLGTS is the linear function of massive polarization tensor, it's impossible to have these two kinds of redundancy. Thus only MLGNS can suffer from EOM and IBP redundancy. In this subsection, we will explain how to get rid of them systematically.

We first discuss how to remove EOM redundancy in amplitude bases.
Although MLGNS $G$ is neutral under massive LG $\otimes_{i=1}^m SU(2)_i$  it is still charged under massless LG $\otimes_{j=m+1}^{N} U(1)_j$. So it must be the polynomial function of massless spinors $\antiket{j}$ or $\ket{j}$ with $j=m+1,\dots,N$, and massive momentum ${p}_{i, \dot \alpha \alpha} \equiv \antiket{i^I}_{\dot\alpha} \bra{i_I}_{\alpha}$ with $i=1,\dots,m$.  As discussed above, the massless amplitude basis can automatically remove the EOM redundancy so it is possible to get rid of it if we first construct the complete set of MLGNT $G$ at its massless limit and then re-construct the original massive $G$ from its massless limit.  Since $G(\antiket{j},\ket{j}, {p}_{i} )$ is massive LG singlet, one $G(\antiket{j},\ket{j}, {p}_{i} )$ will smoothly go to one definite polynomial of massless spinors if all massive momentums go to massless limit,
\bea
{p}_{i, \dot \alpha \alpha}  \to \antiket{i}_{\dot\alpha} \bra{i}_{\alpha} : \, G(\antiket{j},\ket{j}, {p}_{i} ) \to g(\antiket{j},\ket{j},  \antiket{i}\langle i| ),
\eea
where $\antiket{i}_{ \dot \alpha} \bra{i}_{ \alpha}$ is the massless limit of massive momentum ${p}_{i, \alpha \dot \alpha}$ and $g \equiv G (\antiket{j},\ket{j}, \antiket{i}\langle i| )$ is the limit of $G$ when $p_{i, \dot \alpha \alpha}  \to \antiket{i}_{\dot\alpha}\bra{i}_{\alpha}$. We know that the difference of two different $G$ that are related with each other through EOM must be proportional to terms with mass factors so their massless limits must be the same. Reversely the massless spinor polynomial bases $\{g\}$ can be one to one mapped to the independent MLGNS bases $\{G\}$ without EOM redundancy just through replacing massless limit spinors $\antiket{i}$ and $\ket{i}$ in $g$ with the corresponding massive spinors $\antiket{i^I}$ and $\ket{i_{I^\prime}}$ and choosing one pattern of LG index contraction between right-handed spinors $\antiket{i^I}$ and left-handed spinors  $\ket{i_{I'}}$ (equivalent to momentum replacement $ \antiket{i}_{\dot\alpha}  \bra{i}_{\alpha} \to {p}_{i, \dot \alpha \alpha}$). Notice that different LG indices contractions in $g$ produce different $G$ but only one of them is primary and the others are EOM redundant because they have the same massless limit. Based on these discussions, we find that, to construct the complete MLGNS $G$ without EOM redundancy, we should first construct the complete set of its massless limit basis $g$ and then map $g$ to $G$ by restoring the original massive spinors from their massless limit. In the rest of this subsection, we discuss how to construct the complete massless basis $g$ without IBP redundancy.

For a MLGTS $\mathcal{A}_{[\lambda]}^I$ in $SU(2)_r$ representation ${[\lambda]}$, its corresponding MLGNS $G$ should be in the same $SU(2)_r$ representation and it also should be $SU(2)_l$ singlet to preserve Lorentz symmetry, which means that the total number of left-handed spinors should be even  $\sum_{k=1}^N n_k=\mbox{even}\equiv L$, where $n_k$ is the number of massive spinor $\ket{k^I}$ or massless spinor $\ket{k}$. To be massive LG neutral, the number of massive spinor $\antiket{i^I}$ and $\ket{i_I}$ should be equal, and massless LG symmetry requires that the difference between the number of massless spinor $\antiket{j}$ and $\ket{j}$ in $G$ should be equal to the twice of massless particle-$j$ helicity,
\bea \label{eq:spinor_number}
&&\tilde{n}_i-n_i=0\ \ \ \ \,,\ \mbox{with} \; \; i=1,\dots, m \nonumber\\
&&\tilde{n}_j-n_j=2h_j\ ,\ \mbox{with} \; \; j=m+1,\dots, N,
\eea
where $h_j$ is the helicity of particle-$j$ and $\tilde{n}_i$ is the number of right-handed spinor $\antiket{i^I}$ (same for $\tilde{n}_j$).
Its massless limit $g$ should also satisfy these constraints. Since systematical construction of the complete set of massless spinor polynomial $g$ is first proposed in~\cite{Henning:2019enq,Henning:2019mcv}, we will follow the discussion in~\cite{Henning:2019enq,Henning:2019mcv}  to briefly introduce this method. In the massless limit, the LG are just trivial Abelian group $\otimes_{k=1}^{N} U(1)_k$ so we can embed it into a global symmetry $U(N) \supset \otimes_{k=1}^N U(1)_k$ through embedding the massless spinor $\tilde{\lambda}_{\dot\alpha}^k \equiv \antiket{k}$ ($\lambda_{ k \alpha} \equiv  \ket{k}$)  into the (anti-) fundamental representation of $U(N)$ symmetry with $k=1,\dots, N$.
So one polynomial $g$ is a basis of $U(N)$ representation, which corresponds to an SSYT of this $U(N)$ representation, and reversely it can also be written down through this SSYT based on the permutation symmetry of the $U(N)$ indices.
For example, the product of a right/left-handed spinor pair can be obtained from the $2/(N-2)$ rows and $1$ column $U(N)$ SSYT
\bea \label{eq:YD_spinorproduct}
\Yvcentermath1   \young(i,j) &=& ( \tilde{\lambda}_{ \dot\alpha}^i  \tilde{\lambda}_{ \dot\beta}^j -\tilde{\lambda}_{ \dot\alpha}^j  \tilde{\lambda}_{ \dot\beta}^i )  =  [ij]  \nonumber \\
\Yvcentermath1 \!\!\!\!\!\!N-2 \left \{ \begin{array}{c}   \blue{ \begin{Young}
    ${k_1}$  \cr
    ${k_2}$  \cr
    $\cdot$\cr
    $\cdot$\cr
 \end{Young}}
 \end{array}\right. \!\!\!&=&  \frac{\left( \epsilon^{ij k_1 \ldots k_{N-2} } +\mbox{anti-sym in}\; k_1.. k_{N-2} \right)}{(N-2) !} \lambda_{ i \alpha} \lambda_{ j \beta}   \nonumber \\
 &=&  \vev{ i  j}\epsilon^{ij k_1 \ldots k_{N-2} }   \;.
\eea
Notice that the columns in the SSYT associated with the $U(N)$ indices of $\lambda$ are in blue to distinguish with $\tilde{\lambda}$ indices.

Since $g$ can suffer from IBP redundancy, which means that   the spinor polynomials $g$ from some $U(N)$ representations are not kinetic independent. What kind of SSYT is free of IBP?  Next we will briefly discuss how to obtain kinetic independent SSYT. If $g(\tilde{\lambda})$ is holomorphic function of right-handed spinors its $U(N)$ YD is in the same shape as its $SU(2)_r$ YD. Without losing generality, we suppose that its $U(N)$ YD is in the shape of $[r_1,r_2]$ so $g(\tilde{\lambda})$ is holomorphic function of $(r_1+r_2)$ right-handed spinor $\tilde{\lambda}s$, where $[i_1,i_2,...,i_n]$ donates the YD which has $n$ rows and $i_j$ box at $j-$th row. And its YD can be in the following shape,
\eq{ \label{eq:leftYD}
    g(\tilde{\lambda})
    =\Yvcentermath1 \renewcommand\arraystretch{0.05} \setlength\arraycolsep{0.2pt}
    \begin{array}{c} \overbrace{\yng(1)\cdots\yng(1)\cdots\yng(1)}^{r_1}\\
    \underbrace{\yng(1)\cdots\yng(1)}_{r_2} {\color{white} \ \ \ \ \ \ \ \,} \end{array}\,.
}
While, if $g(\lambda)$ is holomorphic function of $L$ left-handed spinor $\lambda s$ and is also Lorentz scalar, its $U(N)$ YD is $N-2$ rows and $L/2$ columns,
\bea \label{eq:rightYD}
    g(\lambda)
    =\Yvcentermath1\underbrace{  \blue{  \begin{array}{ccc}
    \yng(1,1)& \cdots & \yng(1,1) \\
    \vdots &  \ddots & \vdots \\
    \yng(1,1)& \cdots & \yng(1,1)
     \end{array}}  }_{L/2}
\left. \begin{array}{c}
     \\
     \\
     \\
 \end{array}  \!\!\!\!\right \} N-2\,.\!\!\!\!\!\!\!\!\!\!
\eea
For the non-holomorphic case, $g(\lambda,\tilde{\lambda})$ is in the reducible $U(N)$ representation from the out product of the representation of  right- and left-handed spinors in Eq.~(\ref{eq:leftYD}) and (\ref{eq:rightYD}), and it can be decomposed into $U(N)$ irreducible representations via Littlewood-Richardson rules
\bea \label{eq:YD}
    g(\tilde{\lambda},\lambda)\!\!\!\!\!\!\!\!
    &=\Yvcentermath1 N-2 \left \{ \begin{array}{c}
     \\
     \\
     \\
 \end{array}  \right.\!\!\!\!\!
 \underbrace{  \blue{  \begin{array}{ccc}
    \yng(1,1)& \cdots & \yng(1,1) \\
    \vdots &  \ddots & \vdots \\
    \yng(1,1)& \cdots & \yng(1,1)
    \end{array}}  }_{L/2}  \!\times \,
    \renewcommand\arraystretch{0.05} \setlength\arraycolsep{0.2pt}\Yvcentermath1\begin{array}{c} \overbrace{\yng(1)\cdots\yng(1)\cdots\yng(1)}^{r_1}\\
    \underbrace{\yng(1)\cdots\yng(1)}_{r_2} {\color{white} \ \ \ \ \ \ \ \,} \end{array}\nonumber\\
    &=\Yvcentermath1 N-2 \left \{ \begin{array}{c}
     \\
     \\
     \\
 \end{array}  \right.\!\!\!\!\!
     \begin{array}{ccccc }
\blue{ \yng(1,1) }& \blue \cdots &\blue{ \yng(1,1)} \yng(1,1)&  \cdots &\yng(1,1) \\
\blue\vdots &  \blue\ddots & \blue\vdots \quad \quad & & \\
\blue{ \yng(1,1)}& \blue\cdots  & \blue{ \yng(1,1)} {\color{white} \yng(1,1)}  & &
     \end{array}
     \!\!\!\!\!\!\!\!\!\!\!\!\!\!\!\!\!\!\!\!\!\!\!\!\!\!\!\!\!\!\!\!\!\!\!\!\!\!
     \!\!\!\!\!\!\!\!\!\!\!\!\!\!\!\!\!\!\!\!\!\!
     \renewcommand\arraystretch{0.05}\begin{array}{ccccccc}
\color{white}\yng(1) &\color{white}\cdots &\color{white}\yng(2) &\color{white}\cdots &\color{white}\yng(1) &\cdots  &\yng(1)\\
\color{white}\yng(1,1,1,1,1)&  &  &  &  &  &
     \end{array}
     \!\!\!\!\!\oplus  \cdots
\eea
where $\cdots$ represents the other irreducible representations.
It was proved in~\cite{Henning:2019enq,Henning:2019mcv} that only the first irreducible YD does not contain an overall factor of total momentum $P = \sum_{k=1}^{N}  p_k$ so $g(\antiket{k},\ket{k})$ from the first YD, which is obtained by just gluing the blue and white YD simply without shifting around white YD, are primary and complete. Then the complete MLGNS $G(\antiket{j}, \ket{j}, p_i )$ without EOM and IBP redundancy can be obtained from these primary polynomials $g(\antiket{j},\ket{j},\antiket{i}\langle i| )$ by restoring massive LG indices contraction in any way,
\bea
G^{[\eta]}_d(\antiket{j}, \ket{j}, p_i) = g^{[\eta]}_d(\antiket{j},\ket{j},\antiket{i}\langle i|){\vert}_{\antiket{i}\langle i| \to p_i},
\eea
where the superscript $[\eta]$ donates its SSYT of $U(N)$ and the subscript $d$ denotes the mass dimension of $G$. We have demonstrated how to construct the complete bases of $A^I_{[\lambda]}$ and $G^{[\eta]}_d$ through YD method separately. Thus the complete massive amplitude bases can be obtained by contracting the  $A^I_{[\lambda]}$ basis with the corresponding complete $G^{[\eta]}_d$ basis. Is it possible that the bases $A^I_{[\lambda]}\cdot G^{[\eta]}_d$ are redundant after Lorentz index contraction? It can be easily found that the amplitude bases are really complete and independent  without redundancy because of the independence of $A^I_{[\lambda]}$ and $g^{[\eta]}_d$ (rigorous proof can be found in App.~\ref{app:independence}).

\section{Some examples}
\label{sec:example}
In this section we explicitly construct the amplitude bases through YD based on above discussions. We first construct all the 3-point amplitude bases of massive gauge boson $W^+-W^--Z$ and then 4-point basis of $\psi-\psi'-Z-h$ at some definite dimension. The general procedure is that first construct the complete MLGTS via $SU(2)_r$ YDs and then construct their corresponding MLGNS with the same $SU(2)_r$ quantum number to form Lorentz singlet bases.
\subsection{$W^+-W^--Z$ amplitude bases}
 For the 3-pt massive gauge boson bases $W^+W^-Z\partial^n$, MLGTS is the linear function of polarazation tensor of each gauge boson in the triplet representation of $SU(2)_r$ and their independent MLGTS can be obtained through decomposition of the outer product of these three $SU(2)_r$ triplet representations. We can get seven independent tensor structures in the following $SU(2)_r$ YD representations,
\bea \label{eq:three_pt}
&&\Yvcentermath1 \young(11) \times \young(22) \times  \young(33) \nonumber \\
&=&\Yvcentermath1\young(112,233)\oplus\young(1122,33)\oplus\young(1123,23)\oplus\young(1133,22) \nonumber \\
&\oplus& \Yvcentermath1  \young(11223,3)\oplus\young(11233,2)\oplus\young(112233)  \nonumber\\
 &\equiv& \mathcal{A}_{[3,3]}^I\oplus \mathcal{A}_{[(4,2)^1]}^I\oplus \mathcal{A}_{[(4,2)^2]}^I\oplus \mathcal{A}_{[(4,2)^3]}^I \nonumber \\
& &\oplus \mathcal{A}_{[(5,1)^1]}^I\oplus \mathcal{A}_{[(5,1)^2]}^I\oplus \mathcal{A}_{[6]}^I
\eea
Then we can read out the tensor structures based on the permutation symmetry of $SU(2)_r$ indices from above YD (the number $i=1,2,3$ in the YD represent the Lorentz indices of spinor $\antiket{i^I}_{\dot \alpha}$).  Take the first YD as an example, according to the permutation symmetry of their $SU(2)_r$ indices, we can get the expression of MLGTS
\bea
&&\mathcal{A}_{[3,3]}^{\{I_1,I_2\}, \{J_1,J_2 \},\{K_1, K_2\} }
\equiv \Yvcentermath1 \young(112,233)  \nonumber \\
&&=4[{1^{I_1}2^{J_1}}] [{1^{I_2}3^{K_1}}]  [{2^{J_2}3^{K_2}}],
\eea
Its corresponding partners $G$ should be also Lorentz scalar.  For three particles dynamics, $s_{ij}$ is trivial and just a constant function of mass,
\bea
    2s_{ij} = 2 p_i\cdot p_{j}=(\epsilon_{ijk} m_k)^2- m_i^2-m_j^2.
\eea
Since $G$ must be the function of Mandelstam variables $s_{ij}$,  $G$ is just a constant. Meanwhile, we also can not construct a valid SSYT as in Eq.~(\ref{eq:YD}), which means it is not a dynamical polynomial.  So all the amplitude bases with tensor structure $\mathcal{A}_{[3,3]}^{I}$ is just $\mathcal{A}_{[3,3]}^{I}$.

Next we will consider the non-trivial case, such as the bases with the second tensor structure in Eq.~(\ref{eq:three_pt}). This tensor structure is given by
\bea \label{eq:second_YD}
&\mathcal{A}_{[(4,2)^1]}^{\{I_1,I_2\}, \{J_1,J_2 \},\{K_1, K_2\}  }
\equiv \Yvcentermath1 \young(1122,33) \nonumber \\
&=2[{1^{I_1}3^{K_1}}][{1^{I_2}3^{K_2}}]|2^{J_1}]_{\{\dot{\alpha}}
|2^{J_2}]_{\dot{\alpha}'\}} \nonumber \\
&+\;2[{2^{J_1}3^{K_1}}][{2^{J_2}3^{K_2}}]|1^{I_1}]_{\{\dot{\alpha}}
|1^{I_2}]_{\dot{\alpha}'\}} \nonumber \\
&+\;8[{1^{I_1}3^{K_1}}][{2^{J_1}3^{K_2}}]|1^{I_2}]_{\{\dot{\alpha}}
|2^{J_2}]_{\dot{\alpha}'\}}
\eea
To guarantee its partner $G$ to be in the same $SU(2)_r$ quantum number and  LG neutral, it at least contains two right-handed massive spinors to contract with $\mathcal{A}_{[(4,2)^1]}^I$ Lorentz indices and the number of the left-handed massive spinors should be same as right-handed spinors (see Eq.~(\ref{eq:spinor_number})). And thus the lowest dimension of $G^{[\eta]}_d$ is $d=2$, $L=r_1+r_2=2$ (see Eq.~(\ref{eq:leftYD})). The constraints of $G^{[\eta]}_d$ can be satisfied by only one $U(3)$ SSYT which contains $L/2 =1$ column of blue boxes and two columns of white boxes $\begin{array}{c} \blue{ \young(1)}\young(23) \end{array} $. We can write down the amplitude following its permutation symmetry,
\bea
&& G^{[3]}_{d=2}  \equiv \begin{array}{c} \blue{ \young(1)}\young(23) \end{array} \nonumber \\
&=&\Big( \langle i_1i_2\rangle\epsilon^{1i_1i_2}
|2]^{\{\dot{\alpha}}|3]^{\dot{\alpha}'\}}
+\langle i_1i_2\rangle\epsilon^{2i_1i_2}
|3]^{\{\dot{\alpha}}|1]^{\dot{\alpha}'\}}\nonumber\\
&&+\langle i_1i_2\rangle\epsilon^{3i_1i_2}
|1]^{\{\dot{\alpha}}|2]^{\dot{\alpha}'\}} \Big) |_{\antiket{i}\langle i| \to p_i}  \nonumber\\
&=&\langle2_I3_J\rangle |2^I]^{\{\dot{\alpha}}|3^J]^{\dot{\alpha}'\}}
+\langle3_I1_J\rangle |3^I]^{\{\dot{\alpha}}|1^J]^{\dot{\alpha}'\}}\nonumber\\
&&\langle1_I2_J\rangle |1^I]^{\{\dot{\alpha}}|2^J]^{\dot{\alpha}'\}}.
\eea
Notice that after reading out $g$ from $U(3)$ SSYT, $G$ can be got by restoring the LG indices of massive spinors and arbitrarily choosing one LG contraction pattern. In this case the massive LG indices only have one contraction pattern. For higher dimensional polynomials $g^{[\eta]}_{d \ge 4}$, we can not construct a valid YD so higher dimensional $g^{[\eta]}_{d \ge 4}$ can not be primary. This can be seen directly through its dynamics: the extra mass dimension $d-2$ of $G^{[\eta]}_{d \ge 4}$ must come from Mandelstam variables $s_{ij}$ comparing with $G^{[3]}_{d=2}$, which is just mass constant, so $G^{[\eta]}_{d \ge 4}$ is descendent from $G^{[3]}_{d=2}$. In this case, there is only one independent basis totally which is $\mathcal{A}_{[4,2]}^{I} \cdot G^{[3]}_{d=2}$.

For the same reason, the independent partners $G^{[\eta]}_d$ of the rest  tensor structures are also unique so there are total seven bases for 3-pt $W^+-W^--Z$. Following above procedures, we list the lowest dimensional $G^{[\eta]}_d$ for tensor structures,
\bea \label{eq:WWZh}
    \mathcal{A}^I_{[3,3]}&:&G^{\bullet}_{d=0}=\bullet  \nonumber  \\
    \Yvcentermath1 \mathcal{A}_{[(4,2)^{1,2,3}]}^I&:& G^{[3]}_{d=2}= \begin{array}{c} \blue{ \young(1)}\young(23) \end{array}\nonumber  \\
     \mathcal{A}_{[(5,1)^{1,2}]}^I&:& G^{[6]}_{d=4}=\begin{array}{c} \blue{ \young(11)}\young(2233) \end{array}\nonumber  \\
    \mathcal{A}_{[6]}^I&:& G^{[9]}_{d=6}=\begin{array}{c} \blue{ \young(111)}\young(222333) \end{array}.
\eea

\subsection{$\psi-\psi^\prime-Z-h$ amplitude bases}
 Next we construct 4-pt massive amplitude bases of $\psi- \psi^\prime- Z-h$. We'll explicitly show the bases at $d=6$ and $8$, corresponding to the operators $\psi \psi^\prime Zh$ and $\psi \psi^\prime Zh\partial^2$. Following the same procedures as above, the tensor structures can be classified by $SU(2)_r$ representation,
 \bea
    &&\Yvcentermath1  \young(1) \times \young(2) \times \young(33) \times \bullet \nonumber \\
 &=& \Yvcentermath1\young(12,33)\oplus\young(123,3)\oplus\young(133,2)\oplus\young(1233) \nonumber \\
  &\equiv& \mathcal{A}_{[2,2]}^I\oplus \mathcal{A}_{[(3,1)^1]}^I\oplus \mathcal{A}_{[(3,1)^2]}^I\oplus \mathcal{A}_{[4]}^I.
\eea
Since the first tensor structure is Lorentz scalar, the lowest dimensional  basis is just the structure $\mathcal{A}_{[2,2]}^I$ with dimensionless partner $G^\bullet_{d=0} =1$,
\bea
 \Yvcentermath1    \mathcal{A}_{[2,2]}^I G^\bullet_{d=0} =\young(12,33)=[1^{I}3^{\{K_1}][2^{J}3^{K_2\}}].
\eea

The $d=8$ bases $\psi \psi^\prime Zh\partial^2$ should take the structures $\mathcal{A}_{[2,2]}^I$, $\mathcal{A}_{[(3,1)^1]}^I$ and $\mathcal{A}_{[(3,1)^2]}^I$. Imposing the constraints on LG neutral partners,  the $U(4)$ SSYTs of their LG neutral partners can be determined,
\bea
 G^{[(2,2)^{1,2}]}_{d=2} &\equiv& \begin{array}{c} \blue{ \young(1,3)}\young(2,4) \end{array}  \,,\, \begin{array}{c} \blue{ \young(1,2)}\young(3,4) \end{array} \nonumber \\
 G^{[(3,1)^{1,2,3}]}_{d=2} &\equiv& \renewcommand\arraystretch{0.05}  \setlength\arraycolsep{0.2pt} \begin{array}{c} \blue{\young(1)}\young(23) \\ \blue{\young(4)}  {\color{white} \yng(2) } \end{array}\,,\,\begin{array}{c} \blue{\young(1)}\young(24) \\ \blue{\young(3)}  {\color{white} \yng(2) } \end{array}\,,\, \begin{array}{c} \blue{\young(1)}\young(34) \\ \blue{\young(2)}  {\color{white} \yng(2) } \end{array}.
\eea
We now only focus on  $\mathcal{A}^I_{[2,2]}$ partners $ G^{[(2,2)^{1,2}]}_{d=2}$ and the bases with the structures $\mathcal{A}_{[(3,1)^1]}^I$ and $\mathcal{A}_{[(3,1)^2]}^I$ can be found in App.~\ref{app:4_pt}. We can read out $G^{[(2,2)^{1,2}]}_{d=2}$ from their SSYT according to $U(4)$ indices permutation,
\bea
G^{[(2,2)^1]}_{d=2} &=& \Big(\vev{i_1 i_2}\epsilon^{13i_1 i_2}[24] + \vev{i_1 i_2}\epsilon^{23i_1 i_2}[14] \nonumber \\
&&+\vev{i_1 i_2}\epsilon^{14i_1 i_2}[23] + \vev{i_1 i_2}\epsilon^{24i_1 i_2}[13] \Big)|_{\antiket{i}\langle i| \to p_i} \nonumber \\
&=&\langle2_J4_I\rangle[4^I2^J]-\langle1_J4_I\rangle[4^I1^J] \nonumber \\
&&-\langle2_J3_I\rangle[3^I2^J]+\langle1_J3_I\rangle[3^I1^J],
\eea
where in the last identity  we add the massive LG indices and choose one contraction pattern. Following the same procedures, we have
\bea
G^{[(2,2)^2]}_{d=2}&=& \langle3_J4_I\rangle[4^I3^J]-\langle1_J4_I\rangle[4^I1^J]  \nonumber \\
    &-&\langle2_J3_I\rangle[3^I2^J]+\langle1_J2_I\rangle[2^I1^J].
\eea
Then combine them with tensor structure $\mathcal{A}^I_{[2,2]}$ and get the explicit expressions of amplitude bases at $d=8$,
\bea
    &\mathcal{A}^I_{[2,2]} G_{d=2}^{[(2,2)^1]} =8[1^{I}3^{K_1}][2^{J}3^{K_2}]\left(s_{12}+2s_{13}\right) + \mathcal{O}(m_i^2)
    \nonumber \\
    &\mathcal{A}^I_{[2,2]} G_{d=2}^{[(2,2)^2]}
    =8[1^{I}3^{K_1}][2^{J}3^{K_2}] \left(2s_{12}+s_{13}\right) + \mathcal{O}(m_i^2)
\eea
It can be seen easily that these two bases are independent.

\section{EFT operators and Identical particles}
\label{sec:identical}
In this section, we will briefly discuss how to construct bases involving identical particles through YD method. The amplitude bases of identical particles should satisfy boson or fermion statistic: the amplitudes should be symmetric (antisymmetric) under the exchange of the identical bosons (fermions).

 If  the gauge symmetry of massive EFT is $SU(3)_c \otimes SU(2)_L \otimes U(1)_Y$, the complete form of the amplitude bases is the product of gauge group structures and dynamical terms,
\bea \label{eq:total_amplitude}
    \mathcal{M}^I= f_c\,f_W\left(\mathcal{A}_{[\lambda]}^{I} \cdot G^{[\eta]}\right),
\eea
where the $SU(3)_c$ and $SU(2)_L$ gauge symmetry structures $f_c$ and $f_W$ can be systematically constructed by the Young operator~\cite{Li:2020gnx}. Since the total amplitude is a singlet of gauge symmetry, the YD of $SU(3)_c$ ($SU(2)_L$) gauge structures should be singlet in the shape of $[n,n,n]$ ($[m,m]$).
To satisfy bosonic and fermionic statistic, the total amplitude $\mathcal{M}^I$ should be in the $[n_i]$ ($[1^{n_i}]$) representation of the permutation symmetry $S_{n_i}$ associated with $n_i$ identical bosonic (fermionic) particle-$i$.  However, since the amplitude basis is factorized into several parts as in Eq.~(\ref{eq:total_amplitude}), practically we should first construct each part to be in a specific $S_{n_i}$ representation separately via Plethysm operation~\cite{Feng:2007ur,Fonseca:2019yya} and then decompose the inner product of the $S_{n_i}$ representation of each part to get the irreducible $S_{n_i}$ representation of $\mathcal{M}^I$, $[n_i]$ or $[1^{n_i}]$, that satisfies the statistic of identical particles.
Notice that when read out the neutral parts $G$ from the $U(N)$ SSYT, $\frac{L}{2}$ anti-symmetric tensor $\epsilon^{1,2,\ldots,N}$ which are used to raise the $U(N)$ indices of left-handed spinors are included in $G$ (see Eq.~(\ref{eq:YD_spinorproduct})), where $L$ is the total number of left-handed spinors in $G$. So the $S_{n_i}$ symmetry of $U(N)$ SSYT is equal to the inner product of  the $S_{n_i}$ representation of $G$ wave function and these extra tensors $\epsilon^{1,2,\ldots,N}$.  So, at YD level, the needed $S_{n_i}$ YD of $\mathcal{M}^I$ should be equal to the inner product of $[n_i]$ ($[1^{n_i}]$) representation of bosonic (fermionic) identical particles' wave function and $[1^{n_i}]$ representations of these extra tensors $\epsilon^{1,2,\ldots,N}$, which can be expressed as
\bea
  & \bigodot \limits_{a}[\eta_i]_{a}= [1^{n_i}]^{\odot(2s_i+\frac{L}{2})}  \oplus \cdots,
\eea
where symbol $\bigodot$ is inner product, $[\eta_i]_{a}$ is $S_{n_i}$ representation of part-$a$ of amplitude basis $\mathcal{M}^I$, $a =\{f_c, f_W, \mathcal{A}_{[\lambda]}^{I} , G^{[\eta]}\}$, $s_i$ is the spin of identical particle-$i$ and  $[1^{n_i}]^{\odot n}$
represents the inner product of $n$ $[1^{n_i}]$ representation of  $S_{n_i}$.

With this general method based on group theory, we can systematically construct the complete amplitude basis involving identical particles via YD. In App.~\ref{app:id_particle}, we explicitly show how to construct the $5$-pt massive amplitude basis of three vector triplets $V$ and two scalar singlets $\phi$ (more details about dealing with identical particles can also be found in~\cite{Li:2020gnx,Fonseca:2019yya}).

\section{conclusion}
\label{sec:conclusion}
The on-shell SMEFT at the EWSB phase connects with lower energy physics concisely and straightforwardly. It can be directly used in phenomenology calculations without referring to Lagrangian, unlike on-shell massless SMEFT.  To formulate this massive EFT, we first propose a method based on group theory to systematically construct a complete set of massive on-shell amplitude bases corresponding to higher-dimensional operators of massive particles. This method can automatically eliminate redundancies (EOM and IBP) and can be also applicable to construct EFT of any massive particles with any spin.

Different from the amplitude basis of external massless particles, which is charged under abelian LG $U(1)$, the amplitude basis of massive particles is in dim-($2s_i+1$) representation of LG $ SU(2)_i$ associated with massive external particle-$i$ with spin $s_i$.
To guarantee the massive amplitude basis in the proper representation of LG,  we can split it into two parts: MLGTS $\mathcal{A}^I$, which is only charged under massive LG and holomorphic function of massive right-handed spinors, and MLGNS $G$, which is only charged under LG of massless particles. To construct the complete massive amplitude basis, we can separately construct the complete set of the structure $\mathcal{A}^I$ and $G$.

Although different MLGTS $\mathcal{A}^I$ are in the same massive LG representation, they are in different representations of Lorentz subgroup $SU(2)_r$. So tensor structure $\mathcal{A}^I$ can be completely constructed via finding all of its $SU(2)_r$ representations, which are allowed by LG symmetry based on the YD method.

Generally, MLGNS $G$ suffers from EOM and IBP redundancy. To eliminate EOM redundancy, we can first construct the complete set of massless limit structure $g$ of $G$ (EOM of massless spinor is trivial, so massless limit $g$ is automatically free of this redundancy). Then the complete set of $G$ without EOM redundancy can be obtained by replacing the massless limit spinors in $g$ with the original massive ones and arbitrarily choosing one LG contraction pattern.
However, $g$ still suffer from IBP redundancy. To eliminate it, the (left-) right-handed spinors associated with $N$ external particles are embedded in $U(N)$ (anti-) fundamental representation so each $g$ corresponds to a basis of the $U(N)$ representations. The complete set of $g$ free of IBP can be systematically constructed by choosing some specific $U(N)$ representations.

We prove that the massive amplitude bases constructed from the above procedures are indeed independent and complete, guaranteed by group theory. Some examples are given to explicitly demonstrate how to use this method to systematically construct some specific amplitude bases.

This method is also applicable for constructing amplitude bases involving  identical  particles. We briefly discuss about how to deal with identical particles in the framework of this method and give an example to explicitly explain it in detail.

\section*{Acknowledgements}
 J.S. is supported by the National Natural Science Foundation of China under Grants No.12025507, No.11690022, No.11947302; and is supported by the Strategic Priority Research Program and Key Research Program of Frontier Science of the Chinese Academy of Sciences under Grants No.XDB21010200, No.XDB23010000, and No. ZDBS-LY-7003. T.M. is supported by PBC Post-Doctoral Fellowship, the Israel Science Foundation (Grant No.751/19), the United States-Israel Binational Science Foundation~(BSF) (NSF-BSF program Grant No.2018683), and the Azrieli foundation.

\appendix

\section{Independence of amplitude basis $\mathcal{A}^I_{[\lambda]} \cdot G^{[\eta]}$}
\label{app:independence}

An amplitude basis can be factorised into LG tensor and neutral part, $A^I_{[\lambda]}$ and $G^{[\eta]}$.
In this section, we will prove that the bases $A^I_{[\lambda]} \cdot G^{[\eta]}$ are independent.

We have proven that the bases with different massive LG tensor structures must be independent (they can not be related with each other via EOM and IBP) because these bases are independent massive LG tensors. The only possible  redundancy may be from the bases with the same tensor structures. If there are redundancy among these bases, the following identity must be satisfied,
\bea
    \mathcal{A}^I_{[\lambda]} \cdot (\sum_{\eta}Z_{[\eta]} G_d^{[\eta]} + \sum_{i,\eta'}Z_{i[\eta']} m^2_i G_{d-2}^{[\eta']} + \cdots) = 0,
\eea
where $Zs$ are the combination coefficients, and $\cdots$ denotes the higher power terms of mass $m$. Since $\mathcal{A}^I_{[\lambda]}$ is charged under massive LG, each Lorentz component of the LG neutral parts must be zero to satisfy above identity,
\bea
    \sum_{\eta}Z_{[\eta]} G_d^{[\eta]} + \sum_{i,\eta'}Z_{i[\eta']} m^2_i G_{d-2}^{[\eta']} + \cdots = 0.
\eea
Since $G^{[\eta]}_d =g^{[\eta]}_{d} \big{|}_{|i]\langle i| \to p_i}$ can't have an overall factor $m_i^2$ (otherwise its massless limit is null), the first term $\sum_{\eta}Z_{[\eta]} G_d^{[\eta]}$  should be zero and all coefficients of lower dimensional $G_{d^\prime <d}$  should also be zero ($Z_{i[\eta']} =0$,...). Meanwhile this identity $\sum_{\eta}Z_{[\eta]} G_d^{[\eta]}=0$ should also be satisfied at massless limit. Since each polynomial $G_d^{[\eta]}$ is an independent basis of $U(N)$ representations at massless limit, all the coefficient $Z_{[\eta]}$ in this identity must also be zero.  So the amplitude bases $A^I_{[\lambda]}\cdot G^{[\eta]}$ constructed through group  theory are independent.

\section{Amplitude Bases for $\psi -\psi^\prime - Z-h$}
\label{app:4_pt}
In this section, we  write down the explicit expressions of the amplitude bases of $\psi-\psi'-Z-h$ listed in Eq.~(\ref{eq:WWZh}).
\bea
    &&\mathcal{A}^I_{[2,2]}G_{d=2}^{[(2,2)^1]}\nonumber\\
    &\!\!=\!\!&4[1^I3^{\{K_1}][2^J3^{K_2\}}](2m_2^2+m_3^2-m_4^2+2s_{12}+4s_{13})\nonumber\\
    &\!\!\sim\!\!&8[1^I3^{\{K_1}][2^J3^{K_2\}}](s_{12}+2s_{13})
\eea
\bea
    &&\mathcal{A}^I_{[2,2]}G_{d=2}^{[(2,2)^2]}\nonumber\\
    &\!\!=\!\!&4[1^I3^{\{K_1}][2^J3^{K_2\}}](m_2^2+2m_3^2-m_4^2+4s_{12}+2s_{13})\nonumber\\
    &\!\!\sim\!\!&8[1^I3^{\{K_1}][2^J3^{K_2\}}](2s_{12}+s_{13})
\eea
\bea
    &&\mathcal{A}^I_{[(3,1)^1]}\cdot G_{d=2}^{[(3,1)^1]}\nonumber\\
    &\!\!\sim\!\!&-8m_2m_3[1^I3^{\{K_1}]\langle2^J3^{K_2\}}\rangle+8m_1m_3[2^J3^{\{K_1}]\langle1^I3^{K_2\}}\rangle\nonumber\\
    &&-4m_3[1^I2^J]\langle3^{\{K_1}23^{K_2\}}]-4m_3[1^I2^J]\langle3^{\{K_1}13^{K_2\}}]\nonumber\\
    &&+4m_2[1^I3^{\{K_1}]\langle2^J13^{K_2\}}]-4m_1[2^J3^{\{K_1}]\langle1^I23^{K_2\}}]
\eea
\bea
    &&\mathcal{A}^I_{[(3,1)^1]}\cdot G_{d=2}^{[(3,1)^2]}\nonumber\\
    &\!\!\sim\!\!&+8m_2m_3[1^I3^{\{K_1}]\langle2^J3^{K_2\}}\rangle-8m_1m_3[2^J3^{\{K_1}]\langle1^I3^{K_2\}}\rangle\nonumber\\
    &&+4m_3[1^I2^J]\langle3^{\{K_1}23^{K_2\}}]+4m_3[1^I2^J]\langle3^{\{K_1}13^{K_2\}}]\nonumber\\
    &&+12m_2[1^I3^{\{K_1}]\langle2^J13^{K_2\}}]-12m_1[2^J3^{\{K_1}]\langle1^I23^{K_2\}}]
\eea
\bea
    &&\mathcal{A}^I_{[(3,1)^1]}\cdot G_{d=2}^{[(3,1)^3]}\nonumber\\
    &\!\!\sim\!\!&-8m_2m_3[1^I3^{\{K_1}]\langle2^J3^{K_2\}}\rangle-24m_1m_3[2^J3^{\{K_1}]\langle1^I3^{K_2\}}\rangle\nonumber\\
    &&-4m_3[1^I2^J]\langle3^{\{K_1}23^{K_2\}}]+12m_3[1^I2^J]\langle3^{\{K_1}13^{K_2\}}]\nonumber\\
    &&+4m_2[1^I3^{\{K_1}]\langle2^J13^{K_2\}}]-4m_1[2^J3^{\{K_1}]\langle1^I23^{K_2\}}]
\eea
\bea
    &&\mathcal{A}^I_{[(3,1)^2]}\cdot G_{d=2}^{[(3,1)^1]}\nonumber\\
    &\!\!\sim\!\!&+8m_2m_3[1^I3^{\{K_1}]\langle2^J3^{K_2\}}\rangle-8m_1m_3[2^J3^{\{K_1}]\langle1^I3^{K_2\}}\rangle\nonumber\\
    &&+12m_3[1^I2^J]\langle3^{\{K_1}23^{K_2\}}]-4m_3[1^I2^J]\langle3^{\{K_1}13^{K_2\}}]\nonumber\\
    &&+4m_2[1^I3^{\{K_1}]\langle2^J13^{K_2\}}]+12m_1[2^J3^{\{K_1}]\langle1^I23^{K_2\}}]
\eea
\bea
    &&\mathcal{A}^I_{[(3,1)^2]}\cdot G_{d=2}^{[(3,1)^2]}\nonumber\\
    &\!\!\sim\!\!&-8m_2m_3[1^I3^{\{K_1}]\langle2^J3^{K_2\}}\rangle+8m_1m_3[2^J3^{\{K_1}]\langle1^I3^{K_2\}}\rangle\nonumber\\
    &&-12m_3[1^I2^J]\langle3^{\{K_1}23^{K_2\}}]+4m_3[1^I2^J]\langle3^{\{K_1}13^{K_2\}}]\nonumber\\
    &&+12m_2[1^I3^{\{K_1}]\langle2^J13^{K_2\}}]+36m_1[2^J3^{\{K_1}]\langle1^I23^{K_2\}}]
\eea
\bea
    &&\mathcal{A}^I_{[(3,1)^2]}\cdot G_{d=2}^{[(3,1)^2]}\nonumber\\
    &\!\!\sim\!\!&+8m_2m_3[1^I3^{\{K_1}]\langle2^J3^{K_2\}}\rangle+24m_1m_3[2^J3^{\{K_1}]\langle1^I3^{K_2\}}\rangle\nonumber\\
    &&+12m_3[1^I2^J]\langle3^{\{K_1}23^{K_2\}}]+12m_3[1^I2^J]\langle3^{\{K_1}13^{K_2\}}]\nonumber\\
    &&+4m_2[1^I3^{\{K_1}]\langle2^J13^{K_2\}}]+12m_1[2^J3^{\{K_1}]\langle1^I23^{K_2\}}]
\eea
The symbol $\sim$ means that only the independent terms in the amplitude basis expression are written down. Notice that although all $m_i$ are from EOM, the mass factors in some terms is from the EOMs of spinors in MLGTS $\mathcal{A}^I$ which just change the charity of the spinors in $\mathcal{A}^I$ so these terms are generally not EOM redundant.

\section{example for identical particles}
\label{app:id_particle}
In this section, we take the operators $VVV\phi \phi\partial^2$ as an example to explicitly demonstrate how to systematically construct amplitude bases satisfying bosonic and fermionic statistic, where massive vector $V$ and massive scalar $\phi$ is $SU(2)_L$ triplet and singlet. First we need to find all the possible $S_{n_i}$ representations of each part-$a$ whose inner product can contain the needed $S_{n_i}$ representation $[\eta_i] \equiv [1^{n_i}]^{\odot(2s_i+\frac{L}{2})} $ ($\frac{L}{2}=1$, $s_V=1$ and $s_\phi=0$ in this case). Then construct each part-$a$ of the amplitude basis to be in these desired $S_{n_i}$ representations through Plethysm product and finally the amplitude basis in the correct $S_{n_i}$ representation $[\eta_i]$ can be obtained by combining these parts.

The bases contain three identical massive $V$ bosons and two identical massive $\phi$ bosons, so we have $n_V =3$ and $n_\phi =2$. We first consider the permutation symmetry $S_3$  associated with three $V$ bosons. We find that there are three cases that the amplitude basis can be in the correct $S_{3}$ representation $[\eta_V] =[1^3]$,
\bea \label{eq:three_Z}
   &[1^3]_W\odot[3]_{\mathcal{A}}\odot[3]_G=\underline{[1^3]} \nonumber \\
    &[1^3]_W\odot[2,1]_{\mathcal{A}}\odot[2,1]_G=[3]\oplus[2,1]\oplus\underline{[1^3]} \nonumber \\
    &[1^3]_W\odot[1^3]_{\mathcal{A}}\odot[1^3]_G=\underline{[1^3]},
\eea
where $[\eta_V]_{W,\mathcal{A},G}$ is the $S_3$ representation of $SU(2)_L$ gauge structures, MLGTS and MLGNS ($V$ and $\phi$ is neutral under $SU(3)_c$, so their $SU(3)_c$ structures under $S_3$ and $S_2$ are trivial). Notice that  $S_3$ representation of  $SU(2)_L$ gauge structures associated with the three $V$ must be $[1^3]_W$ (corresponding to the $SU(2)_L$ anti-symmetric tensor $\epsilon^{abc}$),  which can be easily seen through Plethysm product.
For two identical scalar $\phi$, the MLGTS is independent of $\phi$ spinors so $S_2$ representation of $\mathcal{A}^I$ is trivial (i.e. $[2]_{\mathcal{A}}$). Thus the MLGNS $G$ must be in $S_2$ representation $[1^2]_G$ to get the correct $S_2$ representation $[\eta_\phi] =[1^2]$ of the amplitude basis,
\bea \label{eq:S2}
    [2]_{\mathcal{A}}\odot[1^2]_G=\underline{[1^2]}.
\eea

Notice that the only nontrivial inner product decomposition is the second identity in Eq.~(\ref{eq:three_Z}). Since the gauge structure is in the totally antisymmetric representation $[1^3]_W$, to get the irreducible representation $[1^3]$ from the inner product decomposition of these three parts, the inner product $[2,1]_{\mathcal{A}}\odot[2,1]_G$ should contain $[3]$. So we need to know the C-G coefficients of  $[3]$ representation generated from the inner product of two $[2,1]$ representations. The C-G coefficient matrix $C_{[3]}$ for $[3]$ representation in the bases of these two $[2,1]_{\mathcal{A},G}$  representations can be expressed as
\bea \label{eq:CG}
 [3] =V_{\mathcal{A}}^T\cdot C_{[3]}\cdot V_{G}\;, \; C_{[3]}=\frac{1}{\sqrt{6}}\left(\begin{array}{cc} 2 &1 \\ 1 & 2  \end{array}\right),
\eea
where $V_{H} = \left\{ [2,1]_{1H}, [2,1]_{2H} \right\}$ is the basis of $[2,1]$ representation and $H =\left \{\mathcal{A}, G \right\}$.

Next we will discuss how to obtain $\mathcal{A}$ in a specific $S_3$ representation via Plethysm operation~\cite{Feng:2007ur,Fonseca:2019yya}. We know that  the massive spinors $\antiket{i^I}$ in $\mathcal{A}$ associated with one identical particle-$i$ with spin $s_i$ are in the $[2s_i]$ $SU(2)_r$ representation. If $\mathcal{A}$ is required to be in the $[\eta_i]_{\mathcal{A}}$ representation of symmetry $S_{n_i}$, the corresponding $SU(2)_r$ YD associated with these $n_i$ identical particles can be obtained by Plethysm product, $[2s_i] \textcircled{p} [\eta_i]_{\mathcal{A}}$. So with requiring $\mathcal{A}$ to be in a particular $S_3$ representation $[\eta_V]_{\mathcal{A}}$, the corresponding $SU(2)_r$ YD associated with three identical $V$ bosons can be obtained by the following Plethysm product decomposition,
\bea \label{eq:Z_Plethysm}
    {[2]}\textcircled{p}[3]_\mathcal{A} &=&[6]\oplus\underline{[4,2]}\oplus[2,2,2] \,, \nonumber \\
    {[2]}\textcircled{p}[2,1]_\mathcal{A}&=&[5,1]\oplus\underline{[4,2]}\oplus[3,2,1] \,, \nonumber \\
    {[2]} \textcircled{p}[1^3]_\mathcal{A}&=&[4,1,1]\oplus\underline{[3,3]}\,.
\eea
Notice that generally the $SU(2)_r$ YD associated with $n_i$ identical particles is a part of the whole YD of $\mathcal{A}$. The whole $SU(2)_r$ YD of $\mathcal{A}$ in a specific $S_{n_i}$ representation is generally obtained by the outer product of the $SU(2)_r$ YD associated with identical particles in Eq.~(\ref{eq:Z_Plethysm}) and the one associated with the other particles.  But in this case, since $\phi$ is scalar, $\mathcal{A}$ only contain the spinors of these three vectors $V$ and thus their $SU(2)_r$ YD is the whole YD of $\mathcal{A}$.
 Since the YDs in the RHS of above equations are $SU(2)_r$ representations, which have at most two rows, the YDs with more than two rows should be ruled out. The derivative operator $\partial^2$ corresponds to two momentum operators in $G$ so dimension of $G$ is $d=2$ and $G$ can provide at most two $SU(2)_r$ indices to contract with $\mathcal{A}$. Thus the YD $[m,n]$ of $\mathcal{A}$ should satisfy the condition $m-n \le 2$ ($m-n$ is the number of un-contracted $SU(2)_r$ indices).
So finally only the underlined YDs in Eq.~(\ref{eq:Z_Plethysm}) are valid and the corresponding wave functions of $\mathcal{A}^I$ can be in the $S_3$ representation $[\eta_V]_{\mathcal{A}}$ at the LHS of these equations. Now we write down all the MLGTSs in the valid $SU(2)_r$ YD $[4,2]$ and $[3,3]$ through decomposing the outer product of the three vector polarization tensors $\Yvcentermath1\young(11)\times\young(22)\times\young(33)$ and we get four valid MLGTSs
\bea \label{eq:four_structure}
     {[4,2]}_1&=& \Yvcentermath1 \young(1122,33) \,, \,\Yvcentermath1 {[4,2]}_2=\young(1123,23)\,, \nonumber \\
    {[4,2]}_3 &=& \Yvcentermath1 \young(1133,22) \,, \,  [3,3]= \Yvcentermath1 \young(112,233)\,.
\eea

There are three different structures $\mathcal{A}$ in the $[4,2]$ representation and the first two decompositions in Eq.~(\ref{eq:Z_Plethysm}) indicate that the $[4,2]$ LG tensor structure in a specific $S_3$ representation should be the combination of these three tensor structures $[4,2]_{1,2,3}$ in Eq.~(\ref{eq:four_structure}). For the degenerate case, the Plethysm product decomposition in Eq.~(\ref{eq:Z_Plethysm}) should be generally parametrised as
\bea \label{eq:general_Plethysm}
    [2] \textcircled{p}[\eta_V]_\mu \supset M_{[\eta]_\mu}^{[\lambda]_\nu} [\lambda]_\nu,
\eea
where $M_{[\eta]_\mu}^{[\lambda]_\nu}$ is the combination coefficient matrix  for getting a tensor $\mathcal{A}$ in the $S_{n_i}$ representation $[\eta_V]_\mu$, $[\lambda]_\nu$ is $SU(2)_r$ representation of the massive spinors associated with three identical vector $V$. In this example, for $[4,2]$ case the tensor structure basis $[\lambda]_\nu=\{ [4,2]_{1,2,3} \}$ and the $S_3$ representation basis of tensor structures is $[\eta_V]_\mu=\{[3],[2,1]_1,[2,1]_2\}$. Since there is only one  MLGTS in $[3,3]$ representation (see Eq.~(\ref{eq:four_structure})), the third decomposition in Eq.~(\ref{eq:Z_Plethysm}) tells us that this tensor structure must be in $[1^3]_{\mathcal{A}}$ representation of $S_3$. After some calculations, the coefficient matrix for $[4,2]$ tensor structures in $S_3$ representation $[\eta_V]_\mu$ is
\bea \label{eq:Plethysm_matrix}
   M_{[\eta]_\mu}^{[\lambda]_\nu} =\left( \begin{array}{ccc}
    1 & 2 & 1\\
    1 & -4 & -2\\
    -2 & -4 & 1
    \end{array} \right).
\eea

Since MLGNS $G$ is the function of spinors associated with both identical vectors $V$ and scalars $\phi$, it should be in a $S_{3} \otimes S_2$ representation ${[\eta_V]}_G \otimes {[\eta_\phi]}_G$, which can be also obtained in a similar way. We first decomposite the Plethysm product of $U(N)$ YD associated with one identical particle-$i$ and the ${[\eta_V]}_G\otimes {[\eta_\phi]}_G$ representation to get the $U(N)$ YD associated with these $n_i$ identical particle-$i$ in representation ${[\eta_V]}_G\otimes {[\eta_\phi]}_G$. Then do the outer product of this YD and the YD associated with other particles to get the whole $U(N)$ YDs of $G$. For example, if $U(5)$ YD of  $G$ is in ${[\eta_V]}_G\otimes {[\eta_\phi]}_G  ={[3]}_G\otimes{[1^2]}_G$ representation, the sub-YD of $U(5)$ SSYT associated with three $V$  and two $\phi$ can be obtained by Plethysm product,
\bea \label{eq:S3S21}
V: [1] \textcircled{p} [3]_G =[3]_V\,, \quad \phi: [1] \textcircled{p} [1^2]_G =[1^2]_\phi\,.
\eea
Notice that sub-YD associated with identical particle-$i$ in above Plethysm product is always in the shape of $[N_i]$  with $N_i =L/2+\tilde{n}_i -n_i$ because the indices of $U(N)$ SSYT of $G$ associated with particle-$i$ are totally symmetric. In this example, since $\tilde{n}_i =n_i$ (see Eq.~(\ref{eq:spinor_number})) and $d=2$ requires that $L=2$, the sub-YD associated with one $V$ or $\phi$ in the Plethysm product is in the shape of $[1]$.  Then the whole $U(5)$ SSYT of $G$ that can contain ${[3]}_G\otimes{[1^2]}_G$ symmetry can be obtained by decomposing the outer product of the two $U(N)$ YDs in Eq.~(\ref{eq:S3S21})
 \bea \label{eq:S3S22}
{[3]}_V \times {[1^2]}_{\phi} &=&  \Yvcentermath1{  \young(123)}_V \times {\young(4,5)}_{\phi} \nonumber \\
& =& \Yvcentermath1  \young(1234,5) \oplus  \renewcommand\arraystretch{0.05}  \setlength\arraycolsep{0.2pt} \begin{array}{c} \blue{\young(1)}\young(23) \\ \blue{\young(4)}  {\color{white}\yng(2)}\\
   \blue{\young(5)}  {\color{white}\yng(2)}\\\end{array},
 \eea
where we label the boxes associated with three $V$ and two $\phi$  by number $1,2,3$ and $4,5$ respectively. Since $G$ is massive LG neutral and $SU(2)_l$ singlet, the first SSYT is ruled out. The first column of the second SSYT is in blue to represent left-handed spinors. At this step, we only get the  YD of G in the $S_5$ representation that contains ${[3]}_G\otimes{[1^2]}_G$ representation but it is now not the eigenstate of ${[3]}_G\otimes{[1^2]}_G$ representation. So we should use the Young operator $\mathcal{Y}_{\scriptsize\young(123)}\otimes \mathcal{Y}_{\scriptsize\young(4,5)}$ of ${[3]}_G\otimes{[1^2]}_G$ representation to act on the $U(5)$ SSYT in Eq.~(\ref{eq:S3S22})  to project out its component in ${[3]}_G\otimes{[1^2]}_G$ representation. Finally the $G$ in  ${[3]}_G\otimes{[1^2]}_G$ representation can be given by
\bea \label{eq:S3S22_1}
G =\Yvcentermath1 \mathcal{Y}_{\scriptsize\young(123)}\, \mathcal{Y}_{\scriptsize\young(4,5)}  \renewcommand\arraystretch{0.05}  \setlength\arraycolsep{0.2pt} \begin{array}{c} \blue{\young(1)}\young(23) \\ \blue{\young(4)}  {\color{white}\yng(2)}\\
   \blue{\young(5)}  {\color{white}\yng(2)}\\\end{array}.
\eea
After constructing the $\mathcal{A}$ and $G$ in $S_3 \otimes S_2$ representation ${[\eta_V]}_{\mathcal{A}} \otimes {[\eta_\phi]}_{\mathcal{A}}$ and ${[\eta_V]}_{G} \otimes {[\eta_\phi]}_{G}$, we can construct the amplitude basis $\mathcal{A}\cdot G$ in the correct representation $[1^3] \otimes [1^2] $  of $S_3 \otimes S_2$  based on above discussions. Notice that since ${[\eta_\phi]}_{\mathcal{A}}$ is always $[2]$, ${[\eta_\phi]}_{G}$ is always required to be $[1^2]$ (see Eq.~(\ref{eq:S2})). If $[\eta_V]_{\mathcal{A}} =[\eta_V]_{G} =[3]$, the corresponding $\mathcal{A}$ can be obtained by substituting the matrix in Eq.~(\ref{eq:Plethysm_matrix}) into Eq.~(\ref{eq:general_Plethysm}),
\bea \label{eq:A3}
    \mathcal{A}_{[3]} &=& M_{[3]}^{[4,2]_\nu}[4,2]_\nu \nonumber  \\
  &=&  \Yvcentermath1 \young(1122,33)+2\ \young(1123,23)+\young(1133,22)\,.\quad\quad
\eea
 The corresponding $G$  can be obtained as in Eq.~(\ref{eq:S3S22})  through Plethysm product,
 \bea
 G^{[3]}_{d=2} &=& \Yvcentermath1 \mathcal{Y}_{\scriptsize\young(123)}\, \mathcal{Y}_{\scriptsize\young(4,5)}\,
    \renewcommand\arraystretch{0.05}  \setlength\arraycolsep{0.2pt} \begin{array}{c} \blue{\young(1)}\young(23) \\ \blue{\young(4)}  {\color{white}\yng(2)}\\
    \blue{\young(5)}  {\color{white}\yng(2)}\\\end{array} \,.
\eea
 Similarly, if $[\eta_V]_{\mathcal{A}} =[\eta_V]_G =[1^3]$, according to the first identity in Eq.~(\ref{eq:Z_Plethysm}) and similar operations in Eq.~(\ref{eq:S3S21}) and~(\ref{eq:S3S22_1}), the corresponding structures are
\bea
    \mathcal{A}_{[1^3]} &=&\Yvcentermath1 [3,3]=\young(112,233)\nonumber\\
    G^{[1^3]}_{d=2} &=& \Yvcentermath1 \mathcal{Y}_{\scriptsize\young(1,2,3)}\, \mathcal{Y}_{\scriptsize\young(4,5)}\,
    \renewcommand\arraystretch{0.05}  \setlength\arraycolsep{0.2pt} \begin{array}{c} \blue{\young(1)}\young(4) \\ \blue{\young(2)}\young(5)\\
   \blue{\young(3)}{\color{white}\yng(1)}\\ \end{array}\,.
\eea
Finally according to the first and third identity in Eq.~(\ref{eq:three_Z}),  if the gauge structure is in $[1^3]$ representation of $S_3$, these two amplitude bases should be in the correct permutation symmetry representation,
\bea
\mathcal{M}=f_{[1^3]_W}\mathcal{A}_{{[\eta_V]}_{\mathcal{A}}}\cdot  G^{{[\eta_V]}_G}\,, \; \mbox{for} \, [\eta_V] = [3],[1^3]
\eea

For the case $[\eta_V]_{\mathcal{A}} =[\eta_V]_{G}= [2,1]$, the corresponding $\mathcal{A}$ can be obtained  as in Eq.~(\ref{eq:A3}),
\bea
    \mathcal{A}_{[2,1]_\mu} &=& \Yvcentermath1 M_{[2,1]_\mu}^{[4,2]_\rho} [4,2]_\rho\nonumber  \\
    &=& \left \{ [4,2]_1-4[4,2]_2-2[4,2]_3,[4,2]_3-4[4,2]_2-2[4,2]_1 \right \}  \nonumber  \\
 G^{[2,1]_\nu}_{d=2}  &=& \Yvcentermath1 \left \{ \,  \Yvcentermath1 \mathcal{Y}_{\scriptsize\young(12,3)}\,\mathcal{Y}_{\scriptsize\young(4,5)}\, \renewcommand\arraystretch{0.05}  \setlength\arraycolsep{0.2pt} \begin{array}{c} \blue{\young(1)}\young(24) \\ \blue{\young(3)}  {\color{white}\yng(2)}\\
   \blue{\young(5)}  {\color{white}\yng(2)}\\\end{array}\;,\;\Yvcentermath1 \mathcal{Y}_{\scriptsize\young(13,2)}\,\mathcal{Y}_{\scriptsize\young(4,5)}\, \renewcommand\arraystretch{0.05}  \setlength\arraycolsep{0.2pt} \begin{array}{c} \blue{\young(1)}\young(34) \\ \blue{\young(2)}  {\color{white}\yng(2)}\\
   \blue{\young(5)}  {\color{white}\yng(2)}\\\end{array}\,\right \} \,.
   \eea
However the decomposition of their inner product is non-trivial (see the second identity in Eq.~(\ref{eq:three_Z})). With the C-G coefficient in Eq.~(\ref{eq:CG}), we can get the amplitude basis with the correct identical particle permutation symmetry
\bea
       \mathcal{M} &=& f_{{[1^3]}_W} C^{\mu\nu}_{[3]}\mathcal{A}_{[2,1]_\mu} \cdot G^{[2,1]_\nu}_{d=2}\,.
\eea

Finally we will discuss how to construct gauge structure in $[1^3]$ representation. The gauge structure of three vector triplet $V$ should be singlet under $SU(2)_L$, which determines that its $SU(2)_L$ YD should be only in the shape of $[3,3]$ (these three triplet vectors have six $SU(2)_L$ indices). This gauge structure must be in the representation $[1^3]$ of $S_3$ (see the third Plethysm product decomposition in Eq.~(\ref{eq:Z_Plethysm})). So gauge structure $f_W$ can be given by the Young operator of YD $[3,3]$ filling with numbers related to $SU(2)_L$ indices
\bea
    f_{[1^3]_W}=\mathcal{Y}_{[1^3]_W} = \mathcal{Y}_{\begin{array}{c} \begin{Young}
    $1$&$\;1'$&$2$\cr
    $\;2'$&$3$&$\;3'$\cr
    \end{Young}
    \end{array}},
\eea
where the number $i$ and $i^\prime$  filling in the YD denote the two $SU(2)_L$ indices of $i$-th $V$ boson ($V$ is in the two indices symmetric representation of $SU(2)_L$). Thus the gauge structure can be easily read from above SSYT according to their indices permutation symmetry if we decompose Young operator into horizontal permutation $\mathcal{P}$ and vertical permutation $\mathcal{Q}$,
\bea
    \mathcal{Y}_{[1^3]_W} =\mathcal{P}_{ \begin{array}{c} \begin{Young}
    $1$&$\;1'$&$2$\cr
    $\;2'$&$3$&$\;3'$\cr
    \end{Young}
    \end{array}}
    \mathcal{Q}_{ \begin{array}{c} \begin{Young}
    $1$&$\;1'$&$2$\cr
    $\;2'$&$3$&$\;3'$\cr
    \end{Young}
    \end{array}}\,.
\eea
Generally the permutation symmetry of $\mathcal{Q}$ can be described by antisymmetric tensor $\epsilon^{ab}$. Replacing $\mathcal{Q}$  by the tensor structure composed by $\epsilon^{ab}$ which has the same permutation symmetry as $\mathcal{Q}$, we can get the needed gauge structure
\bea
    f_{[1^3]_W}
    &=& \mathcal{P}_{ \begin{array}{c} \begin{Young}
    $1$&$\;1'$&$2$\cr
    $\;2'$&$3$&$\;3'$\cr
    \end{Young}
    \end{array}}
    \epsilon^{a_1a_{2'}}\epsilon^{a_{1'}a_{3}}\epsilon^{a_2a_{3'}}\,,
\eea

where $a_i$ and $a_{i'}$ are the $SU(2)_L$ indices of $i$-th $V$ and gauge structure can be obtained after operator $\mathcal{P}$ acts on the indices of these $\epsilon^{ab}s$.


\begin{thebibliography}{99}

\bibitem{Wilson:1970ag}
K.~G.~Wilson,
Phys. Rev. D \textbf{3}, 1818 (1971)
doi:10.1103/PhysRevD.3.1818

\bibitem{tHooft:1979rat}
G.~'t Hooft,
NATO Sci. Ser. B \textbf{59}, 135-157 (1980)
doi:10.1007/978-1-4684-7571-5\_9



\bibitem{Zwicky:1933gu}
F.~Zwicky,
Helv. Phys. Acta \textbf{6}, 110-127 (1933)
doi:10.1007/s10714-008-0707-4

\bibitem{Rubin:1970zza}
V.~C.~Rubin and W.~K.~Ford, Jr.,
Astrophys. J. \textbf{159}, 379-403 (1970)
doi:10.1086/150317

\bibitem{Davis:1968cp}
R.~Davis, Jr., D.~S.~Harmer and K.~C.~Hoffman,
Phys. Rev. Lett. \textbf{20}, 1205-1209 (1968)
doi:10.1103/PhysRevLett.20.1205


\bibitem{Pontecorvo:1957cp}
B.~Pontecorvo,
Sov. Phys. JETP \textbf{6}, 429 (1957)


\bibitem{Lehman:2015via}
  L.~Lehman and A.~Martin,
  Phys.\ Rev.\ D {\bf 91}, 105014 (2015)
  doi:10.1103/PhysRevD.91.105014
  [arXiv:1503.07537 [hep-ph]].

\bibitem{Lehman:2015coa}
  L.~Lehman and A.~Martin,
  JHEP {\bf 1602}, 081 (2016)
  doi:10.1007/JHEP02(2016)081
  [arXiv:1510.00372 [hep-ph]].

\bibitem{Henning:2015alf}
  B.~Henning, X.~Lu, T.~Melia and H.~Murayama,
  JHEP {\bf 1708}, 016 (2017)
  doi:10.1007/JHEP08(2017)016
  [arXiv:1512.03433 [hep-ph]].

\bibitem{Henning:2017fpj}
  B.~Henning, X.~Lu, T.~Melia and H.~Murayama,
  JHEP {\bf 1710}, 199 (2017)
  doi:10.1007/JHEP10(2017)199
  [arXiv:1706.08520 [hep-th]].


\bibitem{Gripaios:2018zrz}
  B.~Gripaios and D.~Sutherland,
  arXiv:1807.07546 [hep-ph].


\bibitem{Bern:2020ikv}
Z.~Bern, J.~Parra-Martinez and E.~Sawyer,
JHEP \textbf{10}, 211 (2020)
doi:10.1007/JHEP10(2020)211
[arXiv:2005.12917 [hep-ph]].

\bibitem{Jiang:2020mhe}
M.~Jiang, T.~Ma and J.~Shu,
[arXiv:2005.10261 [hep-ph]].


\bibitem{EliasMiro:2020tdv}
J.~Elias Mir\'o, J.~Ingoldby and M.~Riembau,
JHEP \textbf{09}, 163 (2020)
doi:10.1007/JHEP09(2020)163
[arXiv:2005.06983 [hep-ph]].


\bibitem{Baratella:2020lzz}
P.~Baratella, C.~Fernandez and A.~Pomarol,
Nucl. Phys. B \textbf{959}, 115155 (2020)
doi:10.1016/j.nuclphysb.2020.115155
[arXiv:2005.07129 [hep-ph]].

\bibitem{Cheung:2015aba}
  C.~Cheung and C.~H.~Shen,
  Phys.\ Rev.\ Lett.\  {\bf 115}, no. 7, 071601 (2015)
  doi:10.1103/PhysRevLett.115.071601
  [arXiv:1505.01844 [hep-ph]].

\bibitem{Jiang:2020sdh}
M.~Jiang, J.~Shu, M.~L.~Xiao and Y.~H.~Zheng,
[arXiv:2001.04481 [hep-ph]].

\bibitem{Cheung:2014dqa}
C.~Cheung, K.~Kampf, J.~Novotny and J.~Trnka,
Phys. Rev. Lett. \textbf{114}, no.22, 221602 (2015)
doi:10.1103/PhysRevLett.114.221602
[arXiv:1412.4095 [hep-th]].

\bibitem{Cheung:2016drk}
C.~Cheung, K.~Kampf, J.~Novotny, C.~H.~Shen and J.~Trnka,
JHEP \textbf{02}, 020 (2017)
doi:10.1007/JHEP02(2017)020
[arXiv:1611.03137 [hep-th]].

\bibitem{Low:2014nga}
I.~Low,
Phys. Rev. D \textbf{91}, no.10, 105017 (2015)
doi:10.1103/PhysRevD.91.105017
[arXiv:1412.2145 [hep-th]].

\bibitem{Low:2014oga}
I.~Low,
Phys. Rev. D \textbf{91}, no.11, 116005 (2015)
doi:10.1103/PhysRevD.91.116005
[arXiv:1412.2146 [hep-ph]].


\bibitem{Elvang:2010jv}
  H.~Elvang, D.~Z.~Freedman and M.~Kiermaier,
  JHEP {\bf 1011}, 016 (2010)
  doi:10.1007/JHEP11(2010)016
  [arXiv:1003.5018 [hep-th]].

\bibitem{Shadmi:2018xan}
  Y.~Shadmi and Y.~Weiss,
  arXiv:1809.09644 [hep-ph].

\bibitem{Ma:2019gtx}
T.~Ma, J.~Shu and M.~L.~Xiao,
[arXiv:1902.06752 [hep-ph]].

\bibitem{Henning:2019enq}
B.~Henning and T.~Melia,
Phys. Rev. D \textbf{100}, no.1, 016015 (2019)
doi:10.1103/PhysRevD.100.016015
[arXiv:1902.06754 [hep-ph]].


\bibitem{Henning:2019mcv}
B.~Henning and T.~Melia,
[arXiv:1902.06747 [hep-th]].

\bibitem{Li:2020gnx}
H.~L.~Li, Z.~Ren, J.~Shu, M.~L.~Xiao, J.~H.~Yu and Y.~H.~Zheng,
[arXiv:2005.00008 [hep-ph]].

\bibitem{Durieux:2019eor}
G.~Durieux, T.~Kitahara, Y.~Shadmi and Y.~Weiss,
JHEP \textbf{01}, 119 (2020)
doi:10.1007/JHEP01(2020)119
[arXiv:1909.10551 [hep-ph]].

\bibitem{Durieux:2020gip}
G.~Durieux, T.~Kitahara, C.~S.~Machado, Y.~Shadmi and Y.~Weiss,
JHEP \textbf{12}, 175 (2020)
doi:10.1007/JHEP12(2020)175
[arXiv:2008.09652 [hep-ph]].

\bibitem{Falkowski:2020fsu}
A.~Falkowski, G.~Isabella and C.~S.~Machado,
[arXiv:2011.05339 [hep-ph]].

\bibitem{Murphy:2020cly}
C.~W.~Murphy,
[arXiv:2012.13291 [hep-ph]].


\bibitem{Witten:2003nn}
  E.~Witten,
  Commun.\ Math.\ Phys.\  {\bf 252}, 189 (2004)
  doi:10.1007/s00220-004-1187-3
  [hep-th/0312171].

\bibitem{Arkani-Hamed:2017jhn}
  N.~Arkani-Hamed, T.~C.~Huang and Y.~t.~Huang,
  arXiv:1709.04891 [hep-th].

\bibitem{Feng:2007ur}
B.~Feng, A.~Hanany and Y.~H.~He,
JHEP \textbf{03}, 090 (2007)
doi:10.1088/1126-6708/2007/03/090
[arXiv:hep-th/0701063 [hep-th]].


\bibitem{Fonseca:2019yya}
R.~M.~Fonseca,
Phys. Rev. D \textbf{101}, no.3, 035040 (2020)
doi:10.1103/PhysRevD.101.035040
[arXiv:1907.12584 [hep-ph]].

\end{thebibliography}
\end{document}